\documentclass[reprint,
showpacs,
showkeys,
nofootinbib,
nobibnotes,
amsmath,fixed
amssymb,
aps, 
superscriptaddress
]{revtex4-2}

\usepackage{array}[=2016-10-06]
\usepackage[english]{babel}
\usepackage[utf8x]{inputenc}
\usepackage[T1]{fontenc}
\usepackage{amsmath}
\usepackage{graphicx}
\usepackage[export]{adjustbox}
\usepackage{comment}

\usepackage[colorlinks=true,linkcolor=blue,citecolor=blue,urlcolor=blue]{hyperref}

\usepackage{epstopdf}
\usepackage{amssymb}
\usepackage{color}
\usepackage{url}
\usepackage[justification=justified]{subcaption}

\usepackage{float}
\restylefloat*{figure}
\usepackage{placeins}

\usepackage{dcolumn}
\usepackage{bm} 
\usepackage{siunitx}
\usepackage{textcomp} 
\usepackage{multirow}
\usepackage{tabularx} 
\usepackage{ragged2e}
\usepackage[overload]{textcase}

\begin{document}

\title{Study of few-electron backgrounds in the LUX-ZEPLIN detector}

\author{D.S.~Akerib}
\affiliation{SLAC National Accelerator Laboratory, Menlo Park, CA 94025-7015, USA}
\affiliation{Kavli Institute for Particle Astrophysics and Cosmology, Stanford University, Stanford, CA  94305-4085 USA}

\author{A.K.~Al Musalhi}
\affiliation{University College London (UCL), Department of Physics and Astronomy, London WC1E 6BT, UK}

\author{F.~Alder}
\affiliation{University College London (UCL), Department of Physics and Astronomy, London WC1E 6BT, UK}

\author{B.J.~Almquist}
\affiliation{Brown University, Department of Physics, Providence, RI 02912-9037, USA}

\author{C.S.~Amarasinghe}
\affiliation{University of California, Santa Barbara, Department of Physics, Santa Barbara, CA 93106-9530, USA}

\author{A.~Ames} \email{dreames@stanford.edu}
\affiliation{SLAC National Accelerator Laboratory, Menlo Park, CA 94025-7015, USA}
\affiliation{Kavli Institute for Particle Astrophysics and Cosmology, Stanford University, Stanford, CA  94305-4085 USA}

\author{T.J.~Anderson}
\affiliation{SLAC National Accelerator Laboratory, Menlo Park, CA 94025-7015, USA}
\affiliation{Kavli Institute for Particle Astrophysics and Cosmology, Stanford University, Stanford, CA  94305-4085 USA}

\author{N.~Angelides}
\affiliation{Imperial College London, Physics Department, Blackett Laboratory, London SW7 2AZ, UK}

\author{H.M.~Ara\'{u}jo}
\affiliation{Imperial College London, Physics Department, Blackett Laboratory, London SW7 2AZ, UK}
\affiliation{STFC Rutherford Appleton Laboratory (RAL), Didcot, OX11 0QX, UK}

\author{J.E.~Armstrong}
\affiliation{University of Maryland, Department of Physics, College Park, MD 20742-4111, USA}

% ...

\begin{abstract}
\noindent
The LUX-ZEPLIN (LZ) experiment aims to detect rare interactions between dark matter particles and xenon. Although the detector is designed to be the most sensitive to GeV/$c^2$--TeV/$c^2$ Weakly Interacting Massive Particles (WIMPs), it is also capable of measuring low-energy ionization signals down to a single electron that may be produced by scatters of sub-GeV/$c^2$ dark matter. 
The major challenge in exploiting this sensitivity is to understand and suppress the ionization background in the few-electron regime. 
We report a characterization of the delayed electron backgrounds following energy depositions in the LZ detector under different detector conditions. In addition, we quantify the probability for photons to be emitted in coincidence with electron emission from the high voltage grids.
We then demonstrate that spontaneous grid electron emission can be identified and rejected with a high efficiency using a coincident photon tag, which provides a tool to improve the sensitivity of future dark matter searches. 
\end{abstract}


\maketitle
\enlargethispage{\baselineskip} 

\section{Introduction}
\label{sec:intro}

The strongest constraints on possible dark matter interactions with atomic matter are set by liquid xenon (LXe) time projection chambers~(TPCs) for dark matter masses of several GeV/$c^2$ and above~\cite{pdg, akerib2022snowmass2021,LZ_WIMPresults}.
A LXe TPC is capable of collecting both scintillation and ionization signals produced by particle interactions. 
Scintillation photons (S1) are produced by the radiative dissociation of xenon dimers and the recombination of electrons with ions promptly following an interaction. Electrons that survive recombination are drifted upward with an electric field toward the liquid surface where they can be extracted into the gas phase. In the gas phase, electroluminescence signals (S2) produced by extracted electrons provide a measurement of the ionization signal strength. The highly localized S2 emission in the gas region can be used to estimate the event's horizontal position, and the time delay between the S1 and S2 signals (electron drift time) measures the depth of the interaction under the LXe surface. Energy partitioning between the S1 and S2 signals is dependent on the particle interaction type and can be used to discriminate between possible dark matter signals and ambient backgrounds~\cite{Xenon10_ERNR}. 

To detect dark matter particles with masses below a few GeV/$c^2$, an experiment must be sensitive to interactions depositing energies at the keV-level and below~\cite{essig2023snowmass2021}. In this energy region, S1 signals from both electron and nuclear recoils are too weak to be observed due to the relatively low light detection efficiency in TPCs (typically around 10\%). However, S2 signals can be collected with high enough efficiency that the extraction of single ionization electron signals may be detected. This allows the energy threshold of a detector to be lowered for ionization-only dark matter searches~\cite{XENON_S2o_limits,DarkSide_S2only_2018,pandax4t_s2o,XENONnT2025_S2o,XENONnT2026_S2Only}, compared to searches requiring both S1 and S2 signals. When combined with the hypothesized Migdal effect~\cite{Migdal_1941,Ibe2017_Migdal}, ionization-only searches using LXe TPC experiments including XENON1T, LUX and PandaX have probed dark matter masses as low as a few MeV~\cite{XENON_Migdal_2019,LUX_Migdal_2021,DM_mediator_PandaX}.

Despite achieving some of the lowest radioactive background levels in any particle detectors~\cite{LZ_backgrounds, LZ_cleanliness, radio_xenonnt, radioactive_pandax}, all LXe TPCs have observed elevated rates of ionization backgrounds. In the energy region relevant to ionization-only dark matter searches, there are few radiation-induced processes that can produce significant backgrounds in the central region of a well-shielded LXe TPC operated deep underground. An exception to this is coherent elastic neutrino nucleus scattering (CE$\nu$NS) by solar neutrinos, but the rates of single-to-few electron pulses observed in current LXe TPC experiments surpass expected rates from CE$\nu$NS by orders of magnitude~\cite{borexino_solarnu,sno_solarnu, PandaX2025_8B,XENONnT2025_8B}. Consequently, this excess is usually attributed to instrumental sources.

These ionization backgrounds have been categorized into three major groups: photoionization by vacuum-ultraviolet  S1 and/or S2 light, delayed electron backgrounds trailing prior energy depositions, and emission from high-voltage electrodes. Photoionization backgrounds were first documented in 2008 by ZEPLIN-II \cite{edwards2008} in which ``unexpected'' single electron (SE) pulses immediately following S1 and S2 pulses were observed. The rate of these pulses appeared to depend on the size of the progenitor pulse; this dependence was later confirmed in ZEPLIN-III \cite{santos2011} and XENON100 \cite{aprile2014}. Both ZEPLIN-III and XENON100 observed photoionization backgrounds up to the maximum drift time, which corresponds to the time required to drift electrons from near the respective TPC cathodes to the liquid surface. XENON100 further showed that the rates of these backgrounds depend strongly on the impurity concentration in the liquid. These conclusions have been supported by more recent studies by LUX \cite{LUX_ebg} and Kopec \textit{et al.} \cite{kopec2021}, but the impurity species responsible for photoionization have not been clearly identified. 

Delayed electron backgrounds have been observed to trail energy depositions at much larger time scales (from milliseconds to seconds) than the maximum drift time in a TPC. 
The majority of these electron pulses have similar horizontal positions to those of the progenitor S2 pulse \cite{LUX_ebg, xenon1t_ebg_2021, akimov_ebg_2016, kopec2021}. With a few exceptions, the delayed electron background rate has been observed to decrease in time in the form of a power-law after a progenitor event \cite{xenon1t_ebg_2021, kopec2021, LUX_ebg,akimov_ebg_2016,Qi2005_LXeSE}. Earlier works \cite{burenkov2009, akimov2012, sorensen2018} suggested that unextracted electrons trapped under the liquid surface may be partially responsible for this phenomenon. However, recent studies from LUX \cite{LUX_ebg}, XENON1T \cite{xenon1t_ebg_2021}, and a single-phase xenon detector~\cite{Qi2005_LXeSE} all support the explanation that liquid bulk impurities may capture drift electrons and release them later on. This question is a subject of study in this work. 

Spontaneous emission of single and multi-electron (ME) pulses from high voltage grid wires is a well-documented effect \cite{tomas2018,linehanDesignProductionHigh2022,abailey}. The most commonly considered sources for this background are wire defects and debris on grid wires. To reduce the chances of debris being deposited on grid wires, extensive cleaning and dust control protocols were implemented during the fabrication and handling of LZ grids \cite{linehanDesignProductionHigh2022}. To counteract the presence of microscopic wire defects, the LZ TPC extraction (gate) grid was passivated with citric acid after fabrication to remove a thin surface layer and form a homogeneous oxide layer. Despite these precautions, intense electron emission from ``hot spots'' was occasionally observed during TPC operation even with otherwise stable high voltage conditions. Fowler-Nordheim (F-N) emission~\cite{Wang:1997ip} has been proposed as a possible mechanism, but observed emission behaviors are not consistent with F-N theory~\cite{abailey}. Namely, emission rates do not monotonically increase with grid voltages, and the electric field on the wire surfaces is not expected to reach the level required for significant F-N emission even with enhancements from defects~\cite{linehanDesignProductionHigh2022}. A hot spot analysis cut was used by LZ to reject data acquired when one or more significant emission centers were active~\cite{LZ_WIMPresults}. This work characterizes the electron emission behavior from these hot spots and also investigates new methods to tag emission pulses so these backgrounds may be rejected. 
High voltage grids can also emit electrons as a result of radioactive decays on the grid wire surfaces, which produce a continuous S2 spectrum up to thousands of electrons~\cite{linehanHighVoltage2022,akerib2026lowenergyradonbackgroundselectrode}. 
However, the low collection efficiency for electrons on the cathode grid surfaces, where most field lines divert away from the drift region, can cause high energy events to appear in the small-S2 region. Additionally, analysis cuts to reject non-bulk events often lose power for small S2s, and as a result, radiogenic grid backgrounds have been observed to contribute significantly to the small-S2 region~\cite{LUX_Migdal_2021,xenon1t_ebg_2021,XENONnT2026_S2Only}.

This paper is organized into four sections. First, Sec.~\ref{sec:detector} describes the LZ detector and the data used in this analysis. Next, Sec.~\ref{sec:delay} characterizes the delayed electron background observed in LZ and examines possible emission mechanisms. Then, Sec.~\ref{sec:grid} studies electron emission from grid surfaces and presents a method to tag and reject such backgrounds using accompanying photon signals. Finally, Sec.~\ref{sec:summary} summarizes the findings of this work and discusses the implications for future low-mass dark matter searches. 

\section{The LZ detector}
\label{sec:detector}

The LZ experiment is located in the Davis Cavern at the Sanford Underground Research Facility (SURF) in Lead, South Dakota. A 4850\,ft rock overburden provides shielding from cosmic ray muons equivalent to 4300\,m of water, reducing their flux by a factor of $3\times10^6$~\cite{Mei_2010,Kudryavtsev_2009}. The detector is composed of multiple nested systems which provide additional background suppression capabilities. At the center is a dual-phase LXe TPC with a 7 tonne active volume. The TPC is instrumented with two arrays of Hamamatsu R11410-20-type 3-inch photomultiplier tubes (PMTs); 253 PMTs are at the top and 241 are at the bottom. The TPC is cylindrical and 145.6\,cm in both diameter and height. The active LXe volume is surrounded by interlocking polytetrafluoroethylene (PTFE) segments to improve light collection efficiency. The electric field required for TPC operation is formed by titanium field-shaping rings embedded between the PTFE segments, and by four wire-mesh electrode grids located at the top and bottom of the TPC. During the first science run (WS2022)~\cite{LZ_WIMPresults}, these electrodes established field values of 193\,V/cm (drift) and 3.9\,kV/cm (extraction, value in the liquid) with a radial variation of approximately 4\% throughout the active volume of the TPC. Under these conditions the maximum drift time for an SE was 951\,$\mu$s and the average SE size was $57.6 \pm 1.9$~photons detected (phd).

The TPC is housed in a double-walled titanium cryostat made from specially-sourced radiopure titanium~\cite{LZ_titanium}. The space between the TPC and inner vacuum vessel constitutes a liquid xenon ``Skin'' detector which is fitted with 93 1-inch and 38 2-inch PMTs. Surrounding the outer titanium vessel are acrylic tanks which comprise the outer detector (OD). The acrylic tanks contain 17 tonnes of gadolinium-loaded (0.1\% by mass) liquid scintillator for neutron detection. Both the Skin detector and the OD function as vetoes for the TPC. Dark matter particles are unlikely to scatter more than once across the three sub-detectors, while TPC-Skin and TPC-OD coincidences are common for gamma and neutron backgrounds. These detection systems are submerged in 238 tonnes of ultra-pure water, housed in a stainless steel tank fitted with 120 8-inch PMTs which detect OD scintillation and water Cherenkov signals. More details on the LZ detector and its construction can be found in Refs.~\cite{LZ_TDR, LZ_detector, LZ_WIMPresults}. 

The LZ experiment started taking science data in late 2021. This analysis uses datasets taken during the commissioning campaign for the high voltage grids prior to WS2022, and some tuning runs at different grid voltages after WS2022. Both datasets were acquired using an 87\,Hz random trigger with 11.1\,ms long event windows, resulting in a near-continuous recording of LZ events with roughly 95\% live time. 
Each recorded event has a 1.1\,ms pre-trigger window, and 10\,ms post-trigger window. An additional 1.1\,ms trigger holdoff was used to prevent events from overlapping. The random trigger setting and the high live time fraction enable us to study the smallest pulses free of trigger efficiency biases. Different electric field configurations were explored during these commissioning runs where drift fields ranged from 110 to 240\,V/cm, and extraction fields ranged from 3.4 to 4.3\,kV/cm. This analysis takes advantage of the varying conditions to study how they affected each component of the electron background. An extraction region-only dataset was also used to study electron emission from the grid surfaces. In this mode, the gate and anode electrodes are held at -4 and +4 kV, while the cathode electrode is set to zero volts, resulting in a WS2022-like extraction region and a reverse drift field region between the gate and the cathode (see~\cite{LZ_TDR} for further details on the electrode configuration). 

This analysis also uses the WS2022 dataset to study the delayed electron emission rates at varying electron lifetime values, and to demonstrate the tagging of grid emission backgrounds. During WS2022 the electron lifetime varied between 5\,ms and 8\,ms, and we opportunistically included data at lifetimes as low as 2\,ms which were excluded from the final WS2022 dataset. The WS2022 data acquisition trigger was configured to record events containing at least one S2 pulse greater than 5 extracted electrons with nearly 100\% efficiency. In addition, a 4\,Hz random trigger and 1\,Hz ``heartbeat'' trigger generated by a Global Positioning System (GPS) clock were injected to record smaller signals including SEs. Each WS2022 event was 4.5\,ms long, with 2\,ms before the trigger and 2.5\,ms after the trigger. A 2.5\,ms trigger holdoff was used to avoid overlapping events. The pre-trigger region of the S2-triggered events can be approximately seen as a random trigger for SE pulses because the S2 trigger efficiency for SE pulses was $<$10\%. 

\section{Delayed electron emission}
\label{sec:delay}

In this section, we characterize the delayed electron background in the LZ detector to study its origin. We also examine the dependence of electron emission rates on detector conditions, and investigate possible delayed release mechanisms. 

For our analysis, we require that progenitors be S2s of at least 200 extracted (``raw'') electrons ($e_{R}$) in size to induce significant delayed electron emission. A drift time cut of $50\,\mu$s to $950\,\mu$s between the S2 and the preceding S1 and a radial cut of 55\,cm are used to select progenitors in the active region of the TPC with well-understood energy and position properties. The drift time cut follows from the fiducial cut implemented in the WS2022 analysis \cite{LZ_WIMPresults}, while the more stringent radial cut excludes progenitors near the circumference of the TPC that may be reconstructed inward due to an S2 light shadowing effect. 

After each progenitor event, all small electron pulses are identified in subsequent data acquisition windows, and the process stops when any pulse larger than 100 $e_{R}$ (``vetoing pulse'') is detected. Progenitors followed by a vetoing pulse within 5\,ms are discarded. We also require a progenitor to be at least 200\,ms after any vetoing pulse to mitigate the aftermath of earlier energy depositions. Small pulses following these disqualified progenitors are not tracked in this analysis. Finally, events with otherwise valid progenitors are rejected if any data buffer for storing PMT digitizations is saturated. This cut effectively excludes progenitor pulses larger than $10^6$~phd and largely mitigates the effects of PMT saturation.

\begin{figure}[t!]
\centering
\includegraphics[width=0.85\linewidth]{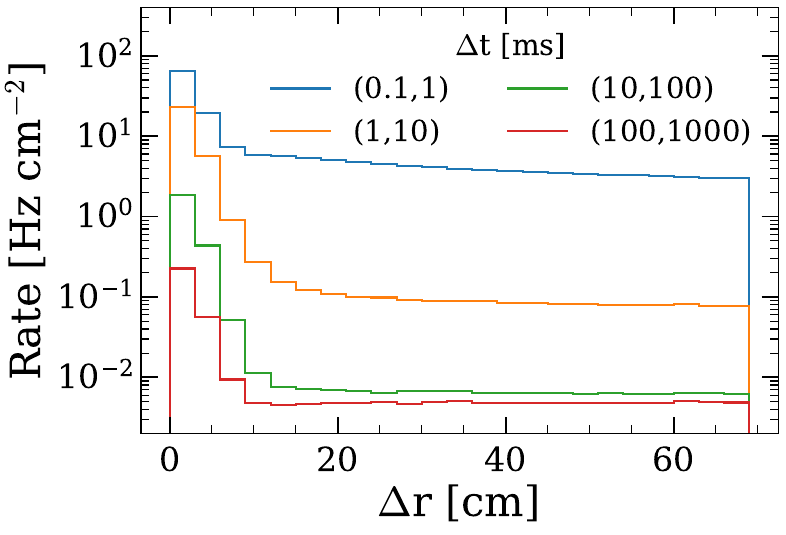}
\caption{Rates of SE pulses at different radial distances relative to their progenitors. These rates were obtained from randomly triggered commissioning datasets. Each curve corresponds to a different time delay window $\Delta t$ following a progenitor. Pulse counts in this plot are normalized by the area of the annular ring subtended by a 3\,cm wide bin in $\Delta r$. This bin width corresponds to the nominal spatial resolution for SE pulses in WS2022. An additional correction (described in the text) is applied to rates at values of $\Delta r$ that would be partially outside of the TPC.}
\label{fig:se_dr}
\end{figure}

As shown in Fig.~\ref{fig:se_dr}, the majority of observed SE pulses share the same horizontal position as the progenitor pulse at both small and large time delays ($\Delta t$). In this analysis we define electron pulses with a radial distance from the progenitor pulse ($\Delta r$) less than 10\,cm as ``position-correlated,'' while the ``position-uncorrelated'' component in this region is estimated using a side band of electron pulses with $\text{20\,cm}< \Delta r < \text{30\,cm}$. Pulses with $\text{10\,cm} < \Delta r < \text{20\,cm}$ are mostly composed of the position-uncorrelated population but have a non-negligible contamination from the position-correlated population. These definitions are used to obtain pulse rates per unit area for each population; i.e., position-correlated rates are normalized by the area of a circle with a radius of 10\,cm around the progenitor position, and the position-uncorrelated rates use the area of a ring with an inner radius of 20\,cm and outer radius of 30\,cm. The normalization also takes into consideration when the circle or ring intercepts the TPC boundary by dividing counts of SE pulses by the fraction of the circumference defined by $\Delta r$ that is within the bounds of the TPC. This correction is very effective, as evidenced by the flat distributions for $\Delta r > \text{30\,cm}$ at large time delays in Fig.~\ref{fig:se_dr}. 

\begin{figure}[t!]
\centering
\includegraphics[width=0.9\linewidth]{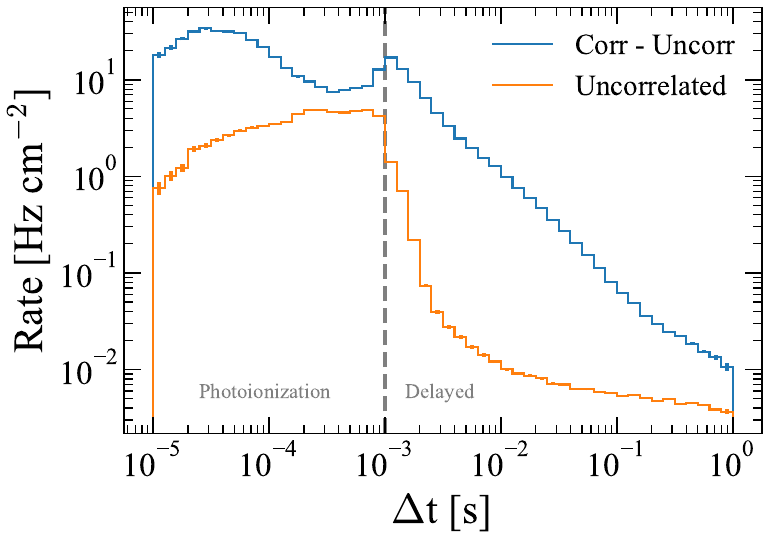}
\caption{Rates of position-correlated (blue) and position-uncorrelated (orange) SE pulses at time $\Delta t$ after progenitor S2 pulses which are at least 200 $e_R$ in size. These rates were obtained from randomly triggered commissioning datasets. The rates are normalized using the method explained in the text, and the position-uncorrelated rates have been subtracted from the position-correlated rates. }
\label{fig:se_dt}
\end{figure}

Examples of the SE rate dependence on delay time $\Delta t$ for both position-correlated and position-uncorrelated electrons are shown in Fig.~\ref{fig:se_dt}. To obtain the live time-normalized electron background rate, we track the detector live time following each progenitor event as a function of $\Delta t$, and then divide the observed pulse counts in each $\Delta t$ bin by the live time in the bin. 

The spectra of both position-correlated and position-uncorrelated electron pulses are dominated by SEs. Immediately following the progenitor S2 and up to the maximum drift time, electrons are produced at a high rate by photoionization of impurities in the LXe and photoelectric effects on metal surfaces~\cite{LUX_ebg, xenon1t_ebg_2021, santos2011, aprile2014}. The high pulse rate compromises both the efficiency of pulse identification and the accuracy of derived pulse quantities. SE pulses in this region often pile up and are misidentified as ME pulses, causing the SE rate to be underestimated. Beyond the maximum drift time, the rate of position-uncorrelated electrons decays very quickly in the first 2--3\,ms. The decay of the rate during this time can be explained as a cascade of the photoionization effect in LXe where electroluminescence from extracted electrons can produce additional photoionization in the liquid. 

After 3\,ms, the pulses are almost exclusively SEs, the rate of which is more than an order of magnitude higher than that of pulses tagged as double electrons. Beyond a time delay of $\sim$10~ms, the rate of pulses identified as ME ($\geq$ 3 electrons) are statistically consistent for the position-correlated and position-uncorrelated distributions, and do not exhibit significant time correlation with progenitors. This suggests that the dominant source of the ME background is independent of preceding energy depositions. The position-correlated SE rate, shown in Fig.~\ref{fig:se_dt}, greatly exceeds that of position-uncorrelated SEs, so we attribute the position-correlated SEs to delayed emission related to earlier progenitor events. Similar to prior studies, the position-correlated SE rate approximately follows a power-law until contributions from position-uncorrelated pulses become significant. Therefore, we subtract the rate of position-uncorrelated pulses from the rate of position-correlated pulses in each $\Delta t$ bin, and fit the resulting distribution with the function $\alpha \Delta t^{\beta}$, where $\alpha$ and $\beta$ are constants, for times between 3 and 200\,ms. At the nominal WS2022 field values we obtain a value of $\beta = -1.13\pm0.01$ with negligible changes in value from varying the choice of fit bounds between 3--200\,ms. When analyzing datasets acquired under the same nominal conditions but at times several months apart, we observe a $\sim$10\% variation in $\beta$, which we attribute to subtle changes in detector conditions and treat as a systematic error for this analysis. The obtained $\beta$ value and uncertainty level are in agreement with the values reported in LUX \cite{LUX_ebg} and XENON1T \cite{xenon1t_ebg_2021}, and will be discussed further in the next subsections.

\subsection{Impurity and drift time dependence}
\label{subsec:impurity}

The total magnitude of the delayed electron rate cannot be analytically evaluated because of the divergence of a power-law integral. Nevertheless, the integrated electron rates in finite delay windows are found to approximately scale with the progenitor S2 size. For this reason, we further normalize the electron rate by the progenitor S2 size to study the rate dependence on detector conditions and progenitor event topology. To help illuminate the origin of these electrons, different S2 normalization factors are considered in this work, including the raw electron number ($e_{R}$, defined in the beginning of Sec. \ref{sec:delay}), the number of electrons reaching the LXe surface ($e_{S}$), the number of electrons produced at the interaction site ($e_{I}$), and the number of electrons lost to impurities during drift ($e_{L}$); the definitions are summarized in Table~\ref{tab:norm}. 

\begin{table}[h!]
    \centering
    \caption{S2 values used in rate normalization. Subscripts in factors are shortened to their first letter for brevity.}
    \begin{tabularx}{\linewidth}{l|X}
         {\bf Factor} [\# of Electrons] & {\bf Description}\\  \hline
        $e_{Raw}$ = $\text{S}2 / \mbox{SE}$& Extracted. "Raw" S2 size divided by SE size.  \\ 
        $e_{Surface}$ = $e_{R} / E_{ee}$& At liquid surface. $e_{R}$ divided by electron extraction efficiency, as calculated from \cite{xuElectronExtractionEfficiency2019} with a linear interpolation.  \\ 
        $e_{Interaction}$ = $e_{S}/e^{-t_{d} /\tau}$& At interaction site. $e_{S}$ divided by drift survival probability. \\ 
        $e_{Lost}$ = $e_{I} - e_{S}$& Lost along drift path.\\
    \end{tabularx}
    \label{tab:norm}
\end{table}

\begin{figure}[t!]
\centering
  \includegraphics[width=1\linewidth]{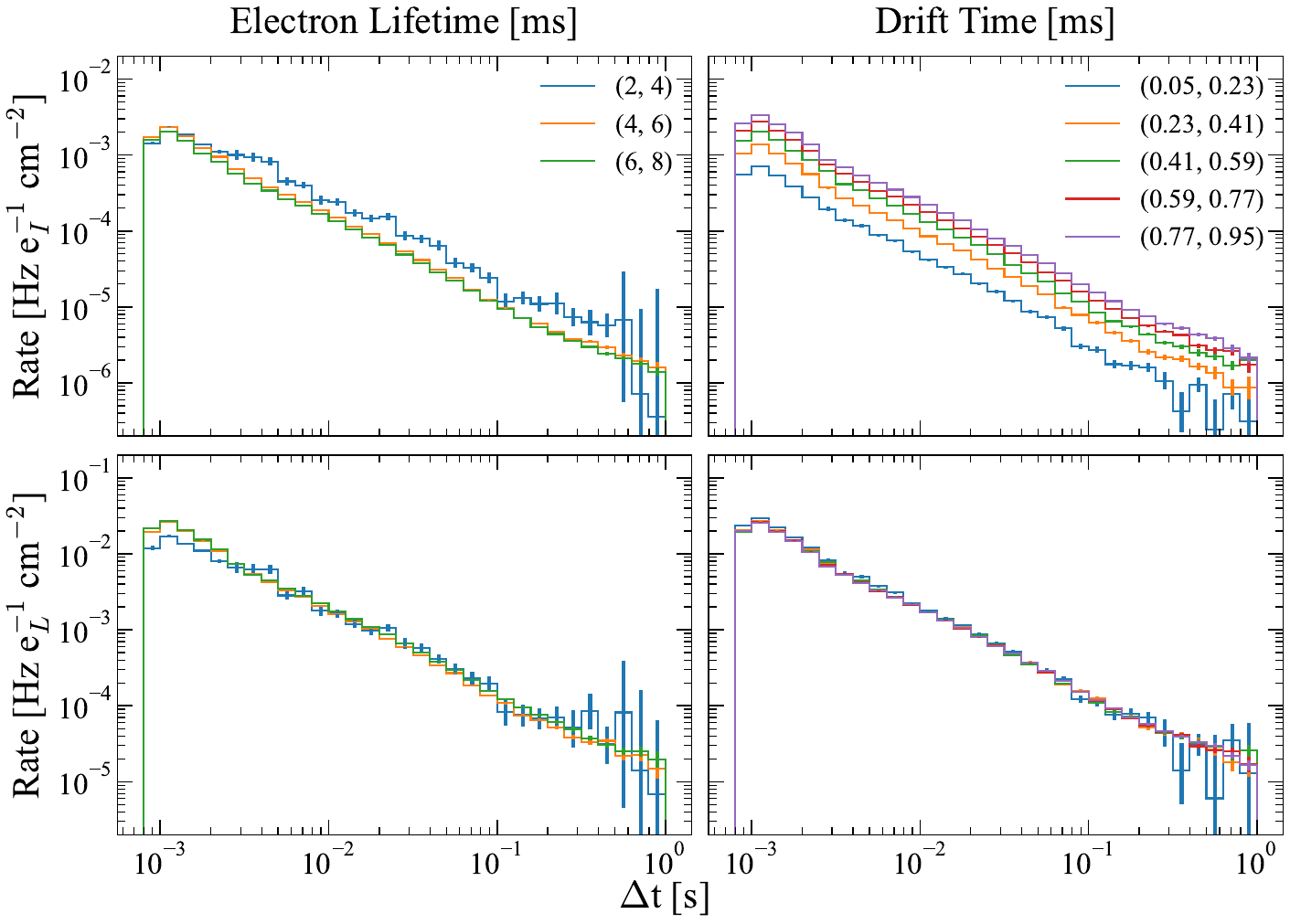}
\caption{Evolution of observed position-correlated electron background rates for electron lifetimes ranging from 2\,ms to 8\,ms (left column) and for progenitor drift times ranging from $50\,\mu$s to $950\,\mu$s (right column). Rates in the top two plots are normalized by $e_I$ while rates in the bottom two plots are normalized by $e_L$; these quantities are defined in Tab. \ref{tab:norm}. Rates in all plots have their corresponding position-uncorrelated rates subtracted. These rates were obtained from the WS2022 dataset.
}
\label{fig:se_dep}
\end{figure}

Figure~\ref{fig:se_dep} shows the rate evolution of the position-correlated electron background at different LXe purity levels and progenitor event depths, using the electron lifetime and S2 drift times as proxies. The upper plots in Fig.~\ref{fig:se_dep} are normalized by $e_{I}$ for progenitors at different depths and impurity concentrations in LXe. For progenitors with the same $e_{I}$ value, longer drift times and higher impurity concentration lead to increased rates of background electrons. Compared to the rate dependence on progenitor drift time, the impurity dependence is relatively weak, especially for high purity levels. We also find that position-uncorrelated background rates do not exhibit a significant dependence on the drift time of the progenitor, but have a similar dependence on the electron lifetime. Therefore, we attribute these backgrounds to delayed emission from earlier progenitor events. 

The lower plots in Fig.~\ref{fig:se_dep} show the same electron rate data, but are normalized instead by $e_{L}$. This normalization nearly eliminates all differences observed in the upper plots, confirming that the drift electrons lost to impurities are primarily responsible for the delayed emission. 

\begin{figure}[t!]
\centering
  \includegraphics[width=0.9\linewidth]{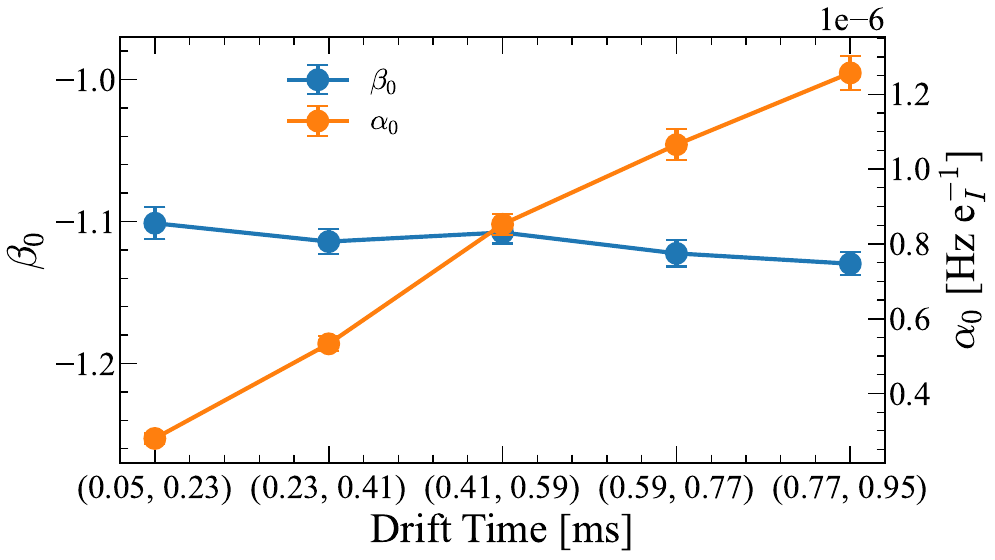}
\caption{Power law fit parameters for $e_I$-normalized rates of position-correlated SEs at different progenitor drift times. The power law amplitude linearly depends on the progenitor drift time, whereas the exponent does not exhibit any significant dependence. Error bars shown here are statistical only and do not include a 10\% systematic uncertainty on values of $\beta$ which is described in the text. }
\label{fig:se_fit_drift}
\end{figure}

Figure~\ref{fig:se_fit_drift} shows the fitted power-law exponents and amplitudes for the electron background rates in the top-right plot of Fig.~\ref{fig:se_dep}. The exponent appears relatively insensitive to the progenitor drift time, but the power-law amplitude increases linearly with drift time. At a typical LZ electron lifetime of 5\,ms with a drift field of 193 V/cm, the integral of position-correlated electrons detected between 3\,ms and 1\,s is approximately 
1.1\% of the progenitor area for a progenitor with a drift time of 0.85\,ms, and is $\sim$0.3\% for a progenitor with a drift time of 0.3\,ms. 
In comparison, 16\% and 6\% of progenitor electrons are captured by impurities in these two scenarios, respectively. Combining the small probability for captured electrons to be released and the relatively weak correlation between the electron rate and the impurity concentration in LXe, we conclude that the impurity species responsible for delayed electron emission is only a subdominant component of electronegative impurities that cause drift electron losses, and that while its concentration mostly tracks that of total impurities, this relationship may break down at low impurity levels. 

This study confirms that impurities in the liquid, which capture drift electrons, are primarily responsible for the delayed electron background, and contributions from un-extracted electrons trapped under the liquid surface are subdominant. 
For a given electron extraction efficiency, the number of electrons not extracted at the surface will be proportional to the number of electrons that survive the drift and arrive at the liquid surface. Therefore, if delayed emission of un-extracted surface electrons could occur, its rate should decrease with electron drift times, which is opposite to observations in Figs.~\ref{fig:se_dep} and \ref{fig:se_fit_drift}. 

This explanation of the position-correlated delayed electron background has also been supported by experimental studies using LUX~\cite{LUX_ebg} and XENON1T~\cite{xenon1t_ebg_2021} data. Because drift electrons are randomly captured by impurities along the drift path, the delayed release of captured electrons should occur independently, producing only delayed SEs. 
In this study, we observed S2 pulses with areas in the double electron (1.5--2.5 SE) region that follow large progenitor S2s, but their dependence on progenitor characteristics and detector conditions is nearly identical to that of SEs discussed above, and their rate is observed as a fixed percentage of the position-correlated SE rate. The same behaviors were also reported from the XENON1T experiment~\cite{xenon1t_ebg_2021}. 
Therefore, we attribute these delayed ``double-electron'' backgrounds to either an upper fluctuation of the SE pulse area or the merge of an SE with its induced electron pulse. Examples of the latter effect include the photoionization of impurities near the liquid surface or photoelectric effects on the gate grids by electroluminescence photons from the initiating SE. 
Unlike XENON1T, position-uncorrelated electron backgrounds in this study are not observed to correlate with progenitor events, possibly a result of the accurate position reconstruction in LZ and our conservative cut choice for position-uncorrelated events. 

\subsection{Electric field dependence}
\label{subsec:field}

Table~\ref{tab:beta-field-table} lists all prior reports of the electron train power law in dual-phase TPCs. An exponent value ranging from -1 to -1.4 is reported across different experiments. Consistent with our analysis shown in Fig.~\ref{fig:se_dep}, the liquid xenon purity, measured as the electron lifetime, does not seem to have significant bearing on the power law slope. 
In addition, these measurements were carried out with varying electron extraction efficiency values, which also do not seem to impact the power-law exponent. In LZ, we analyzed data acquired at a fixed drift field (193\,V/cm) and three different extraction fields: 3410, 3900, and 4300 V/cm. These extraction fields correspond to extraction efficiencies of 50--78\%.  
The power-law exponent values measured at these different field settings were consistent with each other.

Modest changes to the drift electric field also do not seem to significantly alter the power law slope. In LZ, we verified this by analyzing randomly triggered data in the drift region with drift fields at 110\,V/cm, 193\,V/cm and 240\,V/cm with the extraction field at 3900\,V/cm. The power-law exponents for the position-correlated rates at each field setting, with background subtracted, were $-1.04 \pm 0.01$, $-1.13 \pm 0.01$, and $-1.01 \pm 0.01$, respectively, and these results are summarized in Tab.~\ref{tab:beta-field-table}. As mentioned previously, a 10\,\% systematic error is present in this analysis and included in Tab.~\ref{tab:beta-field-table}, which brings these values in agreement with each other. 
 \begin{table}
    \centering
    \footnotesize
    \caption{Reported power law exponents for delayed electron emission at different drift fields. Results in the last 5 rows are from this work. Ref. \cite{kopec2021} reports the $\beta$ value at 500 V/cm and states the values at other fields as ``unchanged''. No fit range is reported in Ref. \cite{LUX_ebg}. Uncertainties in this work and Ref \cite{akimov_ebg_2016} combine statistical and systematic effects. Refs. \cite{LUX_ebg} and \cite{xenon1t_ebg_2021} only report systematic uncertainties while Ref. \cite{kopec2021} only reports the statistical uncertainty.}
    \begin{tabular}{cccccc}
             {\boldmath $E_{drift}$} [V/cm]&  {\boldmath  $\beta$}&  {\boldmath $\tau_{e^-}$} [$\mu$s]& \boldmath {$t_\text{drift}$} \unboldmath [$\mu$s]&   {\bf{Range}} [ms] & \bf{Ref.}\vspace{0.03cm}\\ \hline 100& $-1.20\pm0.04$& 3& 10& 0.03 - 1 ms&\cite{kopec2021}\\
 125& $-1.10 \pm 0.05$& 1000& 750& 2 - 200 &\cite{xenon1t_ebg_2021}\\
 180& $-1.05\pm0.05$& 750& 325& N/A &\cite{LUX_ebg}\\
 200& $-1.20\pm0.04$& 3& 10& 0.03 - 1 &\cite{kopec2021}\\
 500& $-1.20\pm0.04$& 3& 10& 0.03 - 1 &\cite{kopec2021}\\
 1000& $-1.20\pm0.04$& 3& 10& 0.03 - 1 &\cite{kopec2021}\\
 3750& $-1.40\pm0.20$& 9.5& 9& 0.2 - 20 &\cite{akimov_ebg_2016}\\
 110& $-1.04 \pm 0.10$& 9500& 1000 &3 - 200 & -\\
 193& $-1.13 \pm 0.11$& 6000& 1000 &3 - 200 & -\\
 240& $-1.01 \pm 0.10$& 11200& 1000 &3 - 200 & -\\
 3410& $-1.28 \pm 0.29$& 10100& 1000 &3 - 20 & -\\
 3900& $-1.10 \pm 0.29$& 6100& 1000 &3 - 20 & -\\
    \end{tabular}
    \label{tab:beta-field-table}
\end{table}
Notably, Ref.~\cite{akimov_ebg_2016} reported a larger power-law slope at drift fields much higher than in other works. If confirmed, such a field dependence can help test different delayed electron emission hypotheses. Studying this dependence is challenging, however, due to difficulties in sustaining high electric fields in TPCs. To overcome this challenge in LZ, we carried out an analysis using particle interactions in the $\sim$1\,cm liquid volume above the gate in the LZ TPC. This region has fields as high as 3--4\,kV/cm in order to accelerate drift electrons and extract them into the gas. This method enables an in-situ study of the power-law behavior with a drift electric field that is an order of magnitude stronger than the LZ drift region. Despite its shallow depth, the large cross section of the LZ TPC makes the total LXe mass quite significant and enables this study. As discussed in Sec.~\ref{subsec:impurity}, the progenitor drift time only affects the power-law amplitude and has no noticeable effect on the exponent.  

\begin{figure}[t!]
\centering
  \includegraphics[width=1\linewidth]{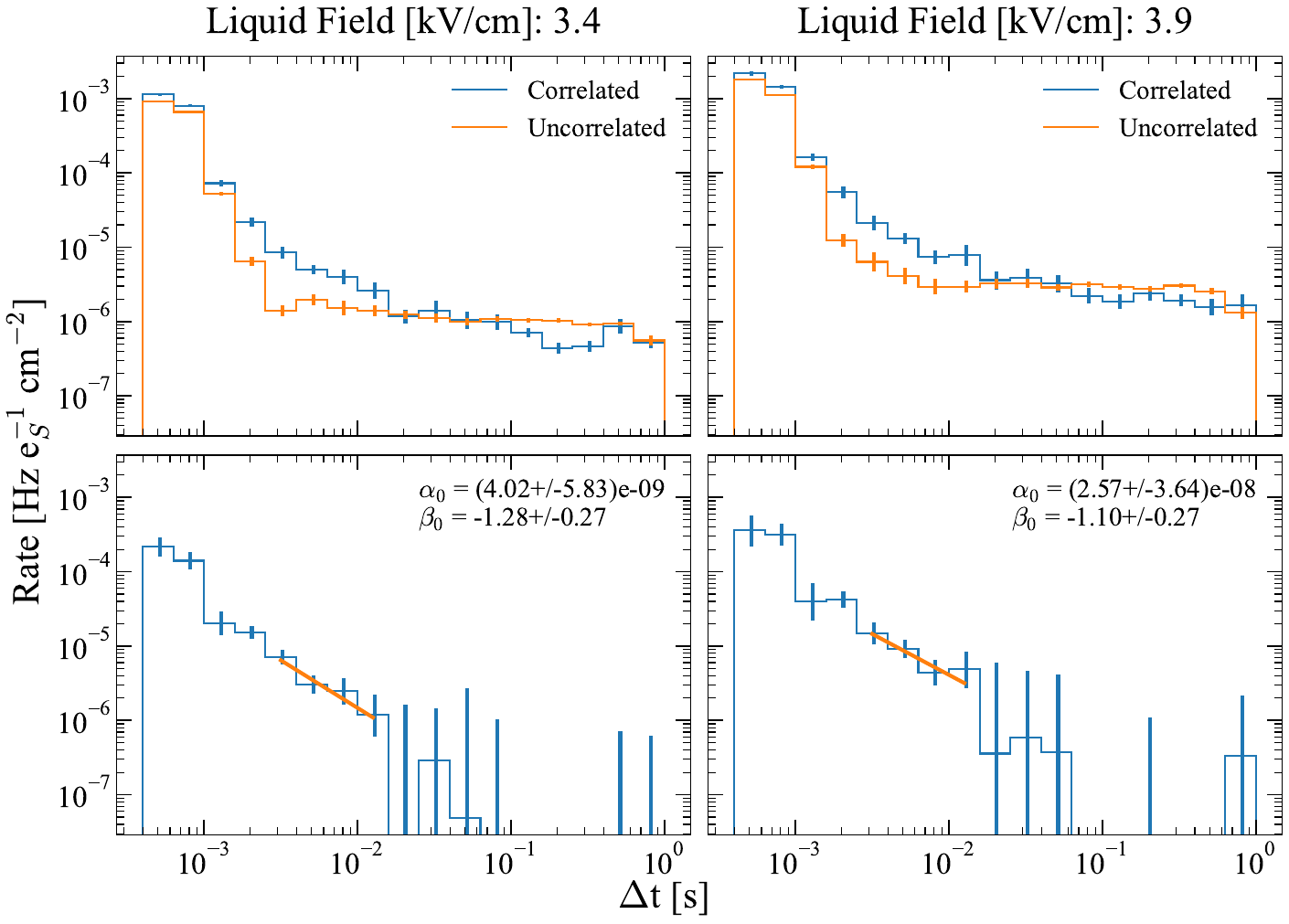}
\caption{Evolution of electron background rates following progenitors which originate in the liquid above the gate grid in LZ, with 3.4 (left) and 3.9\,kV/cm (right) electric fields. These rates were obtained from a combination of the commissioning and grid tuning datasets. The top two plots show both the position-correlated and position-uncorrelated rates. The bottom two plots show power law fits to the remaining position-correlated rates after subtracting the position-uncorrelated rates. While power law fits at each field setting appear consistent with previously observed results, they are subject to large errors. These errors result from a lack of statistics and fit windows that are restricted by un-modeled backgrounds such as the photoionization cascade.
}
\label{fig:se_exfield}
\end{figure}

Figure~\ref{fig:se_exfield} shows the position-correlated electron rates following progenitor events occurring in the extraction region of LZ with two different extraction fields: 3.4 and 3.9\,kV/cm. Although the drift region remains active in these datasets, we select for extraction-region progenitors by requiring a maximum electron drift time of 2.5\,$\mu$s, based on simulations of the LZ extraction field.
In addition, we implicitly require a minimum drift time cut of 1\,$\mu$s; smaller drift times result in merged S1 and progenitor S2 pulses which prevent a drift time from being be calculated. This minimum drift time cut limits this analysis to progenitors $>$1\,mm below the liquid surface and avoids complications from possible interactions of negative ions at the liquid surface; even simple anions such as $O_2^-$ cannot reach the surface and release electrons within the analysis time window at a drift velocity of $\mathcal{O}$(mm/s)~\cite{Hilt1994_IonMobilityLXe}.  

Due to the short drift distance for progenitor ionization signals, very few electrons are captured by impurities in this volume, resulting in low statistics of delayed electrons. However, after subtracting the position-uncorrelated electron rate, we observe clear power-law features up to a 20\,ms delay time window. We fit power laws to the background-subtracted rates at both field settings with a fit range restricted to 3--20\,ms. At 3.4\,kV/cm, we obtain $\beta = -1.28 \pm 0.29$ and at 3.9\,kV/cm we obtain $\beta = -1.10 \pm 0.29$. These results, along with the others reported in this work, are summarized in Tab.~\ref{tab:beta-field-table}.

At some $\Delta t$ values, the uncorrelated rates shown in Fig.~\ref{fig:se_exfield} slightly exceed the position-correlated rates. This is explained by contamination from grid emission electrons (discussed in Sec.~\ref{sec:grid}) in the position-uncorrelated regions, specifically the pre-selected position cut of $\text{20\,cm} <\Delta r<\text{30\,cm}$. Choosing control regions with lower position-uncorrelated rates for background subtraction brings the spectra in Fig.~\ref{fig:se_exfield} (top row) into agreement at large $\Delta t$, but does not significantly alter the obtained exponent values due to the large uncertainties. We also obtain a consistent result from fits with a constant background component added to the power law, in lieu of background subtraction. 

The large statistical and systematic uncertainties in this study prevent us from definitively testing the large power-law exponent at high electric fields like in Ref. \cite{akimov_ebg_2016}. However, with higher statistics being accumulated in LZ and with other tonne-scale LXe experiments being operated across the globe, this method of selecting events only in high electric field regions can enable future studies to investigate the delayed electron background at extraordinarily high fields.  

Unfortunately, there has been no compelling hypothesis for the mechanism behind delayed electron emission. The non-integer power-law exponent observed in LZ and other experiments makes it difficult to attribute this phenomenon to processes such as diffusion~\cite{King1966_diffusion} or recombination~\cite{Kubota1979_recomb}, which can lead to $1/\Delta t$ or $1/\Delta t^2$ time dependence. 
Non-integer power laws are nevertheless observed in systems with energy traps of varying depths~\cite{Huntley2006_PowerLaw}. Taking the example of pumping out a vacuum chamber after it is exposed to air~\cite{Edwards1977_VacuumChamberPowerLaw,Chiggiato2020_outgassing}, when there is only one residual gas species with a specific surface adsorption energy, the chamber pressure will experience an exponential decay when it is pumped on. However, when a range of adsorption energy values are introduced, the overall gas pressure exhibits a power-law time dependence. 
In liquid xenon, Ref.~\cite{sorensen2018} proposes that thermal collisions with neighboring atoms may cause negatively charged impurities to be re-ionized and electrons to be released. In principle, these interactions should also result in an exponential time dependence for the electron emission rate, but the presence of multiple species of impurities with varying electron affinity can lead to an overall emission rate in the form of a power-law of non-integer exponents. Furthermore, the thermal energy of charged particles in an electric field depends on the field strength, so the hypothesis that charged impurities release electrons as a result of thermal collision can be tested by a definitive field-dependence study in the future.

\section{Electron emission from grids}
\label{sec:grid}

\begin{figure}[h]
\centering
\includegraphics[width=0.7\linewidth]{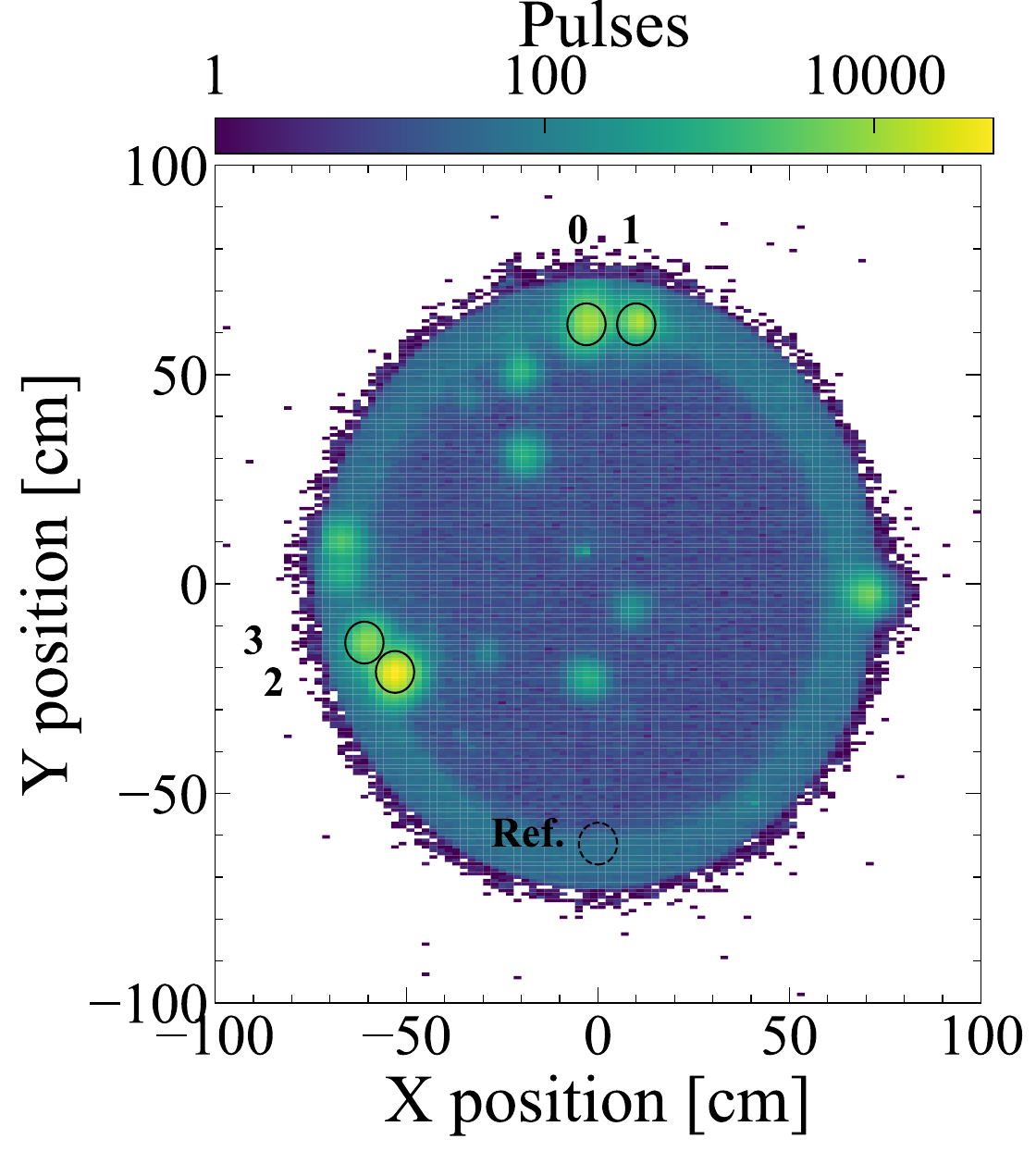}
\includegraphics[width=0.95\linewidth]{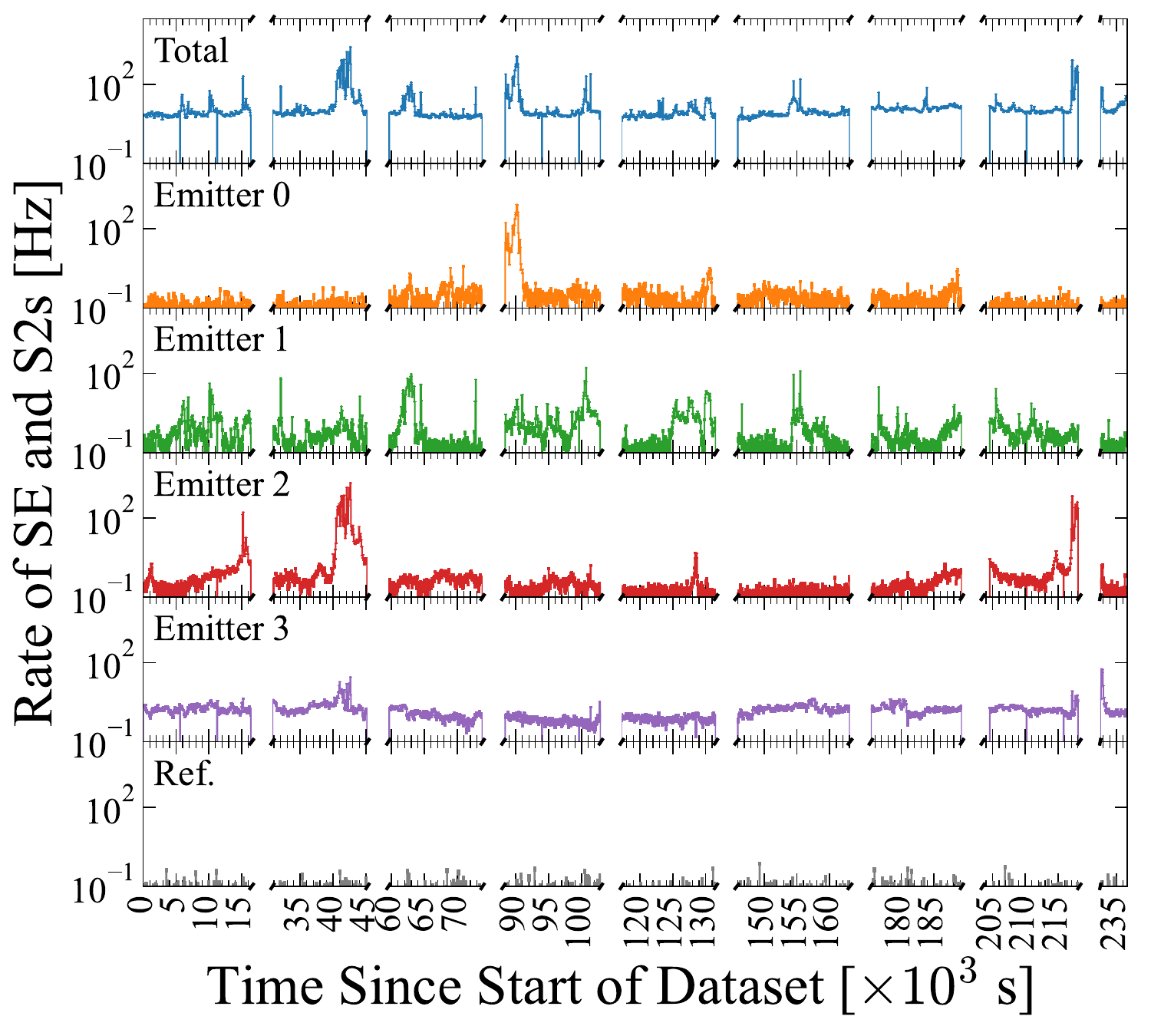}
\caption{Reconstructed X-Y position distribution of S2 pulses (including SEs) in a randomly-triggered extraction region-only dataset (top). The ring feature is due to shadowing effects at the circumference of the TPC that cause S2 positions to be reconstructed inwards. Several X-Y locations, corresponding to active emitters as well as a lower rate reference region, are denoted with black circles and enumerated, and the rates of S2s in those localized regions are shown as a function of time in the bottom panel. Periods with a mixed trigger logic (i.e. not purely random-triggered) are excluded.} \label{fig:xy_pos_se}
\end{figure}
\subsection{Photons accompanying grid electron emission}
\label{sec:grid_characterization}

As discussed in Sec.~\ref{sec:delay}, delayed electron emission following energy depositions in LXe does not appear to contribute significantly to the ME background. In the LUX experiment, ionization backgrounds above SEs are mostly attributed to electric field-induced emission from grid surfaces~\cite{LUX_ebg} and radioactive decays on the grids~\cite{LUX_Migdal_2021}. The field-induced emission could occur even during seemingly stable periods of high voltage operation.

\begin{figure}[b!]
\centering
  \includegraphics[width=0.85\linewidth]{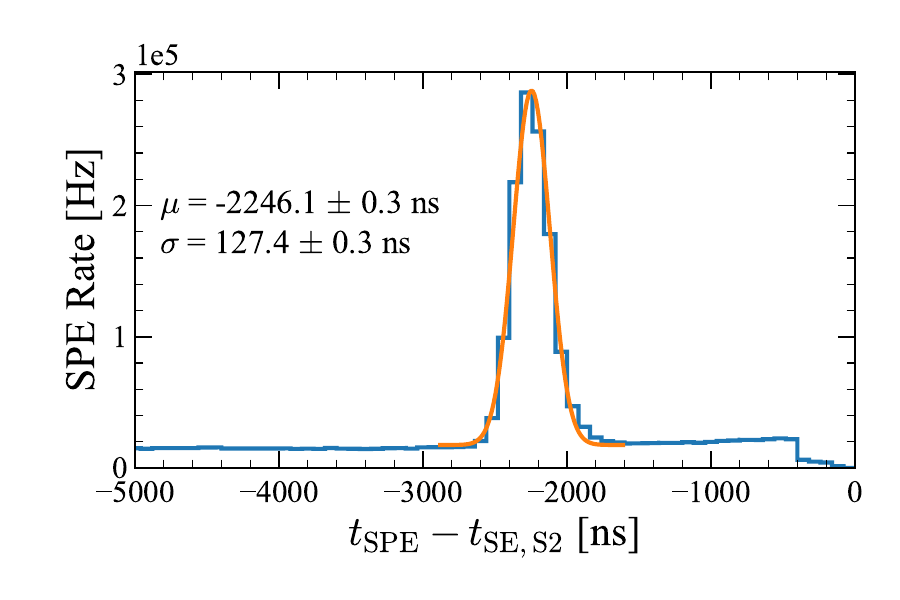}
  \includegraphics[width=0.85\linewidth]{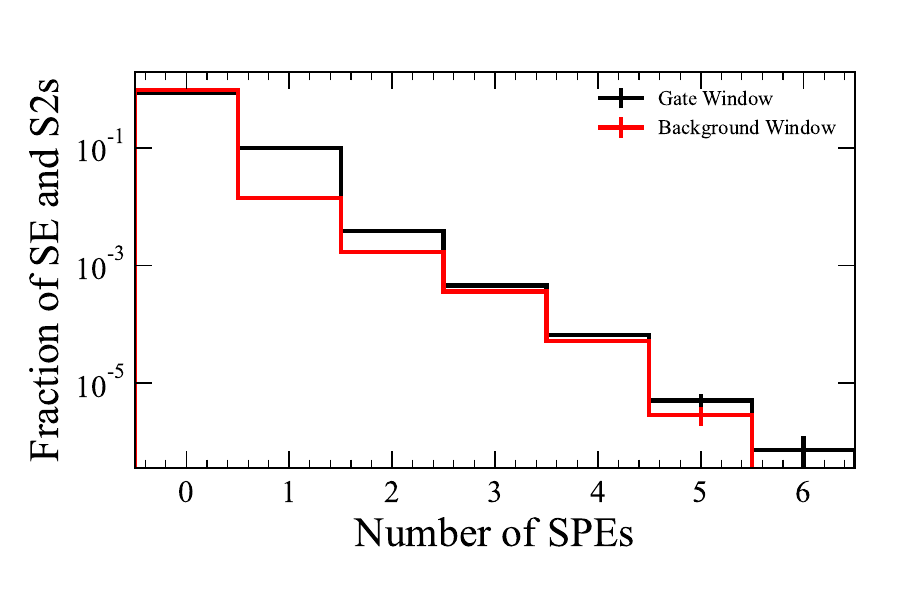}
\caption{Rate of single photo-electrons (SPEs) preceding post-cuts S2 pulses (including SEs), shown in the top panel. The peak position ($\mu$ = -2246.1 $\pm$ 0.3\,ns) is consistent with the drift time associated with the LZ gate grid. The number of SPEs in a time window of $\pm$5$\sigma$ around the gate drift time (black)
and within an equally-sized time window offset from
the gate drift time (red) is shown in the bottom panel.}  \label{fig:spe_peak}
\end{figure}

This section studies electron emission from the LZ gate electrode using an extraction region-only dataset acquired during commissioning. This configuration greatly reduces delayed electron backgrounds following particle interactions in the drift region and is therefore ideal for characterizing grid emission. 
To further reduce the delayed electron background from interactions in the above-gate liquid, 
we veto time periods of $\mathcal{O}(100\,\mathrm{ms})$ following large S2s, where the veto length increases with the size of the S2~\cite{LZ_WIMPresults}. We also require that the S2 pulses do not have S1s larger than 100\,phd within 3\,$\mu$s beforehand to exclude photoionization electrons. 
After applying this selection and additional position reconstruction quality cuts, the X-Y position distribution of the remaining isolated electron pulses is shown in Fig.~\ref{fig:xy_pos_se} (top). Localization of the pulses around specific ``hot spots,'' or spontaneous emitters, is clearly visible. 
In addition to the spatial inhomogeneity, large temporal spikes in the S2 rates are also present in this data set, as shown in Fig.~\ref{fig:xy_pos_se} (bottom). Because the cathode is inactive, the hot spots are attributed to electron emission from defects or debris on the gate electrode. 
During the enhanced electron emission periods, high rates of photons are also observed, a phenomenon also reported in LUX~\cite{abailey}. 
Figure~\ref{fig:spe_peak} (top) shows the arrival time of photons relative to that of S2s, or the rate of single photo-electrons (SPEs) per S2 per second, as a function of the time difference between the SPE and S2. SPEs are defined in the LZ reconstruction algorithm as S1-like signals with light in only one PMT. A prominent peak is observed at $\sim$2.2\,$\mu$s, which is consistent with the drift time from the gate wire to the liquid surface at the WS2022 field. 
As described in Sec.~\ref{subsec:field}, the maximum drift time for extraction region events is approximately 2.5\,$\mu$s, but the drift time for gate-emitted electrons also depends on the exact emission position on the wire and the local field structure.
This observation confirms that these photons and electrons are emitted around the same time and that the electrons originate from the gate grid. 
Similar gate photon-electron coincidences are also observed in a study of radiogenic backgrounds on LZ grids~\cite{linehanHighVoltage2022}; the rate of grid radioactive decays, however, is over an order of magnitude lower than that observed for field-induced emission in the few-electron region and thus its contribution to this analysis is neglected.

The spectrum of photon signals that fall within the gate drift time window ($\pm 5\sigma$ around the fitted peak) is shown in the bottom plot of Fig.~\ref{fig:spe_peak}, 
and compared to that of an equally-sized time window offset from the peak. The photons in the offset time window are not associated with gate electron emission and are treated as a background. 
By taking the difference of these two distributions, we calculate that $8.87 \pm 0.02$\% of all S2 pulses in this data set have an associated SPE preceding them.
This probability increases to $9.15 \pm 0.03$\% when we restrict our event selection to grid emission periods with S2 rates above $\sim$100 Hz in Fig.~\ref{fig:xy_pos_se}. 
With a photon detection efficiency of 0.094 $\pm$ 0.004 phd/photon measured for calibration events near the LZ gate, we estimate that the majority ($64 \pm 2$\%) of gate-emitted electron pulses have an associated photon signal;   
this value is obtained based on a Monte Carlo study that simply assumes a fixed-rate Poisson distribution for the photon emission process. When the light-shadowing effect of the stainless steel grid wires is taken into consideration, the on-grid light collection efficiency should be lower, suggesting that the actual fraction of gate-emitted electron pulses with an associated photon signal should be even higher.

\begin{figure}[t!]
\centering
  \includegraphics[width=0.85\linewidth]{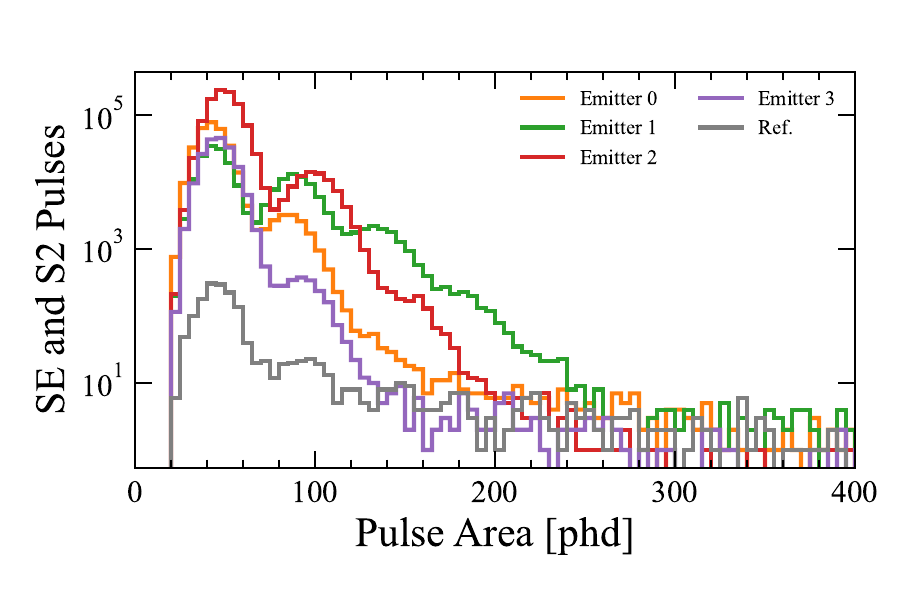}
  \includegraphics[width=0.85\linewidth]{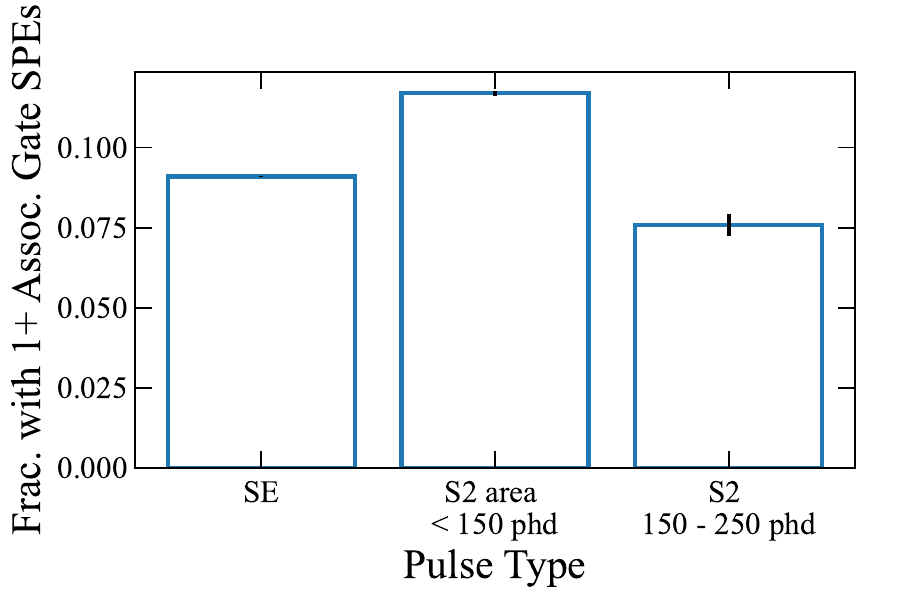}
\caption{S2 pulse area spectra from both identified emitters and a reference region enumerated in Fig.~\ref{fig:xy_pos_se}, shown in the top panel. The bottom panel depicts the fraction of S2s of different pulse sizes with at least one correlated SPE, after subtracting the background from the uncorrelated SPE rate.} \label{fig:spe_tagging}
\end{figure}

Notably, different grid emitters in LZ exhibit different temporal behaviors (Fig.~\ref{fig:xy_pos_se}) and also produce differing pulse area spectra, as shown in the top plot of Fig.~\ref{fig:spe_tagging}, where Emitter 1 emits multiple electrons at a higher rate. However, the number of electrons observed per pulse does not appear to correlate strongly with the probability of coincident photon emission, as shown in the bottom plot of Fig.~\ref{fig:spe_tagging}. There was also no clear dependence seen in the LUX data~\cite{abailey}, which points to some decoupling between the mechanism behind multiple electron emission and photon emission.

\subsection{Implications for dark matter searches}

The vast majority of grid-emitted S2 pulses observed in LZ (Fig.~\ref{fig:spe_tagging}) contain fewer than 5 electrons ($\sim$300\,phd). 
The standard WIMP search of LZ uses an S2 threshold of approximately 10~electrons~\cite{LZ_WIMPresults}, so grid emission only contaminates the region of interest when rates are high enough to pile up significantly.
In WS2022~\cite{LZ_WIMPresults}, we implemented a rate-based hot spot exclusion cut by removing time intervals (120~seconds) where the average rate of low-energy S2s (including SEs) exceeded a certain threshold. 
This cut was demonstrated to be sufficient in reducing spurious S2 backgrounds for a sensitive search of medium to high-mass dark matter candidates.

Searches for low-mass dark matter interactions can greatly benefit from a lower energy threshold than 10~electrons in S2. Grid emission, however, produces spurious S2 pulses below 5--6~electrons, and significantly contaminates the data and weakens the dark matter sensitivity. 
In this section, we demonstrate that the photon signals accompanying grid electron emission can be used to efficiently tag and reject grid electron backgrounds. 
Although we estimate that $>$64\% of LZ gate grid electrons are accompanied by photon signals, the photon detection efficiency in LZ only allows $\sim$9\% of all gate electrons to be tagged using a time coincidence. 
Nevertheless, because grid emission electrons tend to cluster in time and concentrate at certain X-Y locations, tagging one grid electron can allow multiple background electrons to be rejected. 
As a result, a $\gg$9\% grid background rejection efficiency may be achieved. 

\begin{figure}[t!]
\centering
\includegraphics[width=\columnwidth]{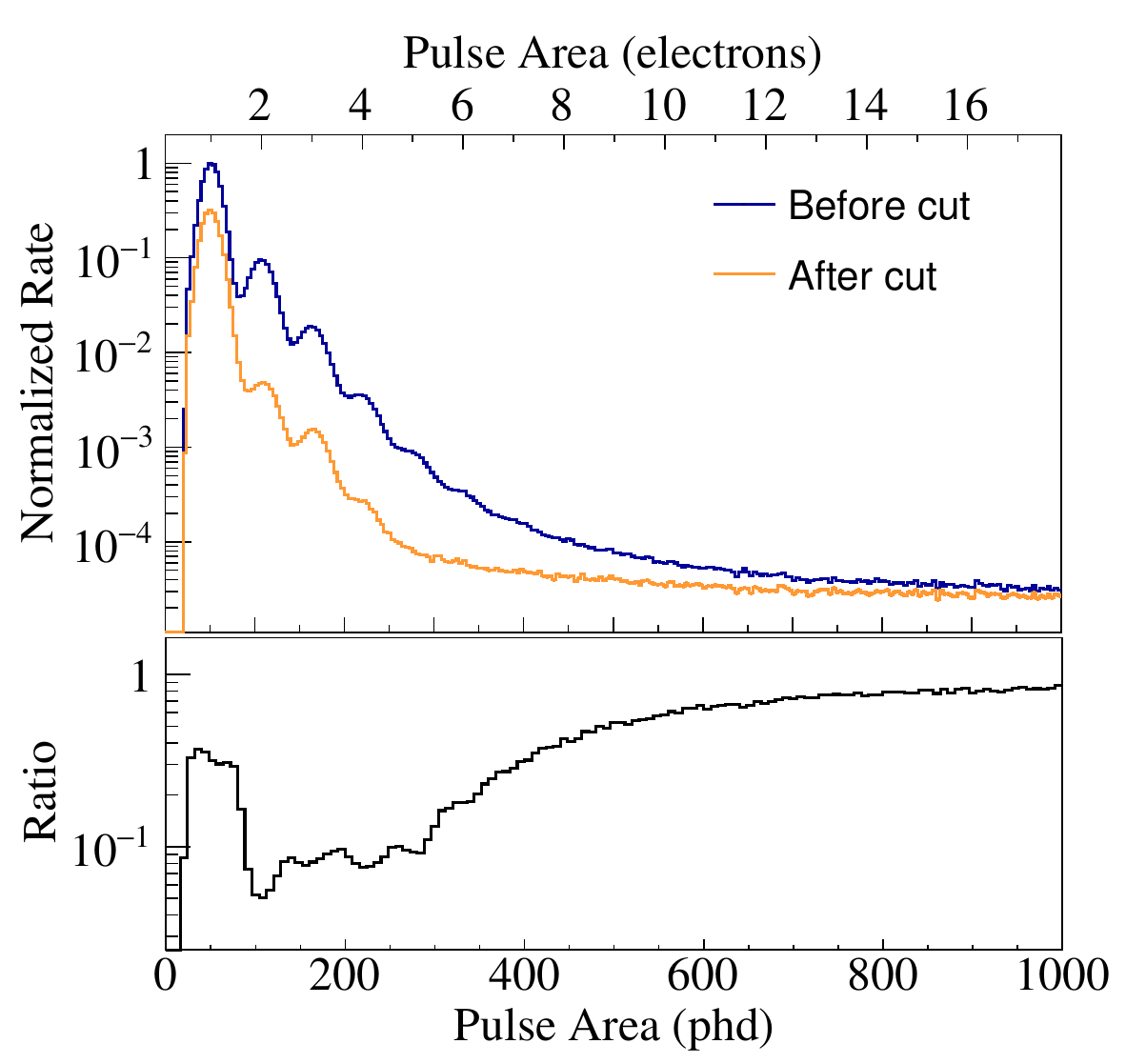}
\caption{ 
Normalized event spectra of S2 pulses identified in a test WS2022 dataset, before (blue) and after (orange) the gate photon emission cut; the ratio between the two is shown in the lower panel.
Cuts applied in this analysis are described in the text; their efficiencies have not been fully evaluated.
} 
\label{fig:pulse_area_sr1}
\end{figure}

As a proof of principle, we tested a photon-based gate emission cut on a fraction of the WS2022 dataset. This dataset is first processed to suppress delayed electron backgrounds and photoionization, as described in Sec.~\ref{sec:grid_characterization}.  An additional pulse quality cut was introduced to suppress misreconstructed S2 pulses. 
We then tag small S2 pulses ($<$300\,phd, including SEs) that are preceded by at least one SPE within the gate drift window. For simplicity, we only exploit the temporal correlation of grid emission electrons in this test by removing all pulses within the event acquisition window (4.5\,ms) if it contains at least one SPE-tagged S2 pulse. 
Figure~\ref{fig:pulse_area_sr1} shows the S2 pulse area spectra before and after the gate photon emission cut. This simple cut suppresses the rate of S2 pulses in the 2--4 electron region, which is dominated by hot spot emission, by over an order of magnitude, and that of SEs by more than a factor of 2. 
On the contrary, large S2s, which are unlikely to be from grid field emission, are less affected by this cut. 
The pulses removed by this cut exhibit a localized spatial and temporal distribution and are consistent with field-induced grid emission.

For low-mass dark matter searches, we usually focus on events with no identified S1s\footnote{The S1 definition is the same as that used in WS2022 where at least 3 photomultiplier tubes must observe light.} within the 4.5~ms event window, and this cut is found to accept $93.60\pm0.26\%$ of S2-only events between 10 and 50 electrons. 
The signal loss is dominated by false SPE tagging, which we further investigate with a different method. 
The WS2022 data, with the TPC drift region active, contains a higher SPE rate than that in the extraction region-only dataset shown in Fig.~\ref{fig:spe_peak}. 
As a result, the probability for a SPE to be randomly paired with a S2 pulse can be significant and cause non-grid S2 pulses to be falsely rejected. 
We carried out a mock analysis by tagging SPEs arriving in a control drift time window that is equal in size but offset from the gate cut interval. 
Then we remove all S2 pulses in an event if any S2 pulse of $<$300\,phd has a tagged SPE in the control coincidence window. 
The SPE-S2 pairs tagged in this way are only from random coincidence and the resulting event rate reduction is a measure of signal loss from random photon coincidence. 
When this mock cut is applied to S2-only events, we found a survival fraction of $97.27\pm0.16\%$ for events with S2 pulses above 300 phd (and below 2,880\,phd, or 50 electrons), and a lower fraction of $90.61\pm0.04\%$ for events with S2 pulses below 300\,phd. 
The survival fractions are different for S2 pulses above and below 300\,phd because a large S2 can only be rejected in this cut if it shares the event window with a tagged S2 of $<$300\,phd, but a small S2 can be additionally rejected if it has a preceding SPE pulse. 
Above 300\,phd, there is minimal dependence on pulse area; the survival fraction for S2-only events above 10 electrons is $98.02 \pm 0.16\%$.

The uncertainties quoted above are statistical only, and a full acceptance analysis is not attempted here because it would depend on data trigger efficiency and complex timing structures of SPE and S2 rates. 
The obtained efficiency values, nevertheless, suggest that
a substantial reduction in grid backgrounds can be achieved with 
a high signal acceptance of $\sim$90\% or more.
It is worth noting that this proof-of-principle analysis does not use the standard LZ hot spot cut that excludes events recorded during periods of high average S2 rates. 
There is considerable overlap between the background populations removed by these two cuts.
However, the event-level method developed in this work can enhance sensitivity to weak grid-emission with an average rate too low to be detected by rate-based cuts. 
This cut can be further optimized to balance background rejection power and signal loss by 
tuning the width of the pulse rejection window and the minimum number of photon-tagged S2s within the window. One can further exploit the position distribution of tagged S2s to further reduce accidental coincidences and to improve the signal efficiency. 
The optimization of this background rejection method can be explored in future work.

\section{Summary}
\label{sec:summary}
The excessive electron backgrounds observed in current LXe TPC detectors must be substantially suppressed for these experiments to improve their sensitivities in low-mass dark matter searches. 
In this work, we thoroughly characterize the delayed electron emission and the spontaneous grid emission in the LZ detector. 
By studying the correlation between delayed electron emission rates and detector conditions, such as the liquid impurity and the drift field applied to the LXe volume, we confirm that drift electrons lost to impurities in LXe are a major contributor to delayed electron emission. 
Spurious electrons from grid emission contribute significantly to few-electron backgrounds in the LZ detector. Using extraction region-only data we find that these electrons are often accompanied by photon emission, confirming observations in LUX.
We further demonstrate that grid-emitted electrons can be efficiently tagged by looking for electron-photon coincidence with a time delay, which substantially reduces the background below 6 electrons in a test WS2022 dataset. Future ionization-only dark matter searches can optimize the photon tagging cut to improve its signal acceptance and background rejection power in the few-electron region. 

\begin{acknowledgments}

The research supporting this work took place in part at the Sanford Underground Research Facility (SURF) in Lead, South Dakota. Funding for this work is supported by the U.S. Department of Energy, Office of Science, Office of High Energy Physics under Contract Numbers DE-AC02-05CH11231, DE-SC0020216, DE-SC0012704, DE-SC0010010, DE-AC02-07CH11359, DE-SC0015910, DE-SC0014223, DE-SC0010813, DE-SC0009999, DE-NA0003180, DE-SC0011702, DE-SC0010072, DE-SC0006605, DE-SC0008475, DE-SC0019193, DE-FG02-10ER46709, UW PRJ82AJ, DE-SC0013542, DE-AC02-76SF00515, DE-SC0018982, DE-SC0019066, DE-SC0015535, DE-SC0019319, DE-SC0025629, DE-SC0024114, DE-AC52-07NA27344, \& DE-SC0012447. This research was also supported by U.S. National Science Foundation (NSF); the UKRI’s Science \& Technology Facilities Council under award numbers ST/W000490/1, ST/W000482/1, ST/W000636/1, ST/W000466/1, ST/W000628/1, ST/W000555/1, ST/W000547/1, ST/W00058X/1, ST/X508263/1, ST/V506862/1, ST/X508561/1, ST/V507040/1 , ST/W507787/1, ST/R003181/1, ST/R003181/2, ST/W507957/1, ST/X005984/1, ST/X006050/1; Portuguese Foundation for Science and Technology (FCT) under award numbers PTDC/FIS-PAR/2831/2020; the Institute for Basic Science, Korea (budget number IBS-R016-D1); the Swiss National Science Foundation (SNSF) under award number 10001549. This research was supported by the Australian Government through the Australian Research Council Centre of Excellence for Dark Matter Particle Physics under award number CE200100008. We acknowledge additional support from the UK Science \& Technology Facilities Council (STFC) for PhD studentships and the STFC Boulby Underground Laboratory in the U.K., the GridPP [1,2] and IRIS Collaborations, in particular at Imperial College London and additional support by the University College London (UCL) Cosmoparticle Initiative, and the University of Zurich. We acknowledge additional support from the Center for the Fundamental Physics of the Universe, Brown University. K.T. Lesko acknowledges the support of Brasenose College and Oxford University. The LZ Collaboration acknowledges the key contributions of Dr. Sidney Cahn, Yale University, in the production of calibration sources. This research used resources of the National Energy Research Scientific Computing Center, a DOE Office of Science User Facility supported by the Office of Science of the U.S. Department of Energy under Contract No. DE-AC02-05CH11231. We gratefully acknowledge support from GitLab through its GitLab for Education Program. The University of Edinburgh is a charitable body, registered in Scotland, with the registration number SC005336. The assistance of SURF and its personnel in providing physical access and general logistical and technical support is acknowledged. We acknowledge the South Dakota Governor's office, the South Dakota Community Foundation, the South Dakota State University Foundation, and the University of South Dakota Foundation for use of xenon. We also acknowledge the University of Alabama for providing xenon. For the purpose of open access, the authors have applied a Creative Commons Attribution (CC BY) license to any Author Accepted Manuscript version arising from this submission. Finally, we respectfully acknowledge that we are on the traditional land of Indigenous American peoples and honor their rich cultural heritage and enduring contributions. Their deep connection to this land and their resilience and wisdom continue to inspire and enrich our community. We commit to learning from and supporting their effort as original stewards of this land and to preserve their cultures and rights for a more inclusive and sustainable future.

\end{acknowledgments}

\newpage

\bibliographystyle{apsrev4-2}
\bibliography{reference}

\end{document}